\documentclass[fleqn,11pt,twocolumn]{wlscirep}

\usepackage{textcomp}
\usepackage{amssymb,amsfonts,amsmath}
\usepackage{upgreek}
\usepackage{upgreek}
\usepackage{multirow}
\usepackage{rotating}

\title{When Hearing Defers to Touch}

\author[1]{Charles Hudin}
\author[2]{Vincent Hayward}

\affil[1]{CEA, LIST, Sensorial and Ambient Interfaces Laboratory, 91191, Gif-sur-Yvette, France}
\affil[2]{Sorbonne Universit\'e, Institut des Syst\`emes Intelligents et de Robotique, 75005, Paris, France}

\begin{abstract}\textbf{
Hearing is often believed to be more sensitive than touch. This assertion is based on a comparison of sensitivities to weak stimuli. The respective stimuli, however, are not easily comparable since hearing is gauged using acoustic pressure and touch using skin displacement. We show that under reasonable assumptions the auditory and tactile detection thresholds can be  reconciled on a level playing field. The results indicate that the capacity of touch and hearing to detect weak stimuli varies according to the size of a sensed object as well as to the frequency of its oscillations. In particular, touch is found to be more effective than hearing at detecting small and slow objects.}\\[6pt]
\end{abstract}

\begin{document}

\flushbottom
\maketitle

\thispagestyle{empty}

\section*{Introduction}


Despite obvious differences in their respective perceptual acumen, hearing and touch share the common purpose of detecting the rapid movements of external objects. To do so, touch detects the vibrations of the skin caused by contact with external objects and hearing detects the airborne waves radiated by distant objects. 

The surfaces that we tactually interrogate often emit sounds and remote oscillating objects can also come in direct contact with the skin. This is the case, for instance, when a flying insect lands on your arm or when you grab a buzzing alarm clock. Tactile and auditory stimuli are thus frequently correlated because hearing and touch access common mechanical sources.

The coupling between hearing and touch is also mediated by the body itself since certain acoustic waves can be felt and certain bodily vibrations can be heard~\cite{Barany-38,Corso-63,Hakansson-et-al-84,Elsayed-et-al-15}. Research has shown that hearing and touch frequently participate jointly in the elaboration of percepts, resulting in numerous neural correlates of co-activation of somatosensory and auditory areas~\cite{Foxe-et-al-00,Foxe-et-al-02,Fu-et-al-03,Murray-et-al-05,Kayser-et-al-05,Cappe-et-al-05,Schurmann-et-al-06,Caetano-Jousmaki-06,Hackett-et-al-07,Yau-et-al-09,Serino-et-al-11,van-Wassenhove-Schroeder-12,Ro-et-al-13}. There are also myriads of perceptual interactions between hearing and touch~\cite{Von-Bekesy-59,Gescheider-et-al-74,Lederman-79,Jousmaki-Hari-98,Caclin-et-al-02,Calvert-Spence-Stein-04,Schurmann-et-al-04,Bresciani-et-al-05,Tajadura-Jimenez-et-al-08,Alais-et-al-10,Occelli-et-al-11,Canzoneri-et-al-12,Lin-Makio-12,Ito-Ostry-12,Okazaki-et-al-12,Frissen-et-al-12,Desloge-et-al-14,Landry-et-al-14,Deroy-et-al-16}. 

Tactile sensitivity ranges from continuous skin loading to vibrations of 1~kHz, while audition responds to acoustic pressure from 20~Hz to 20~kHz, thus, hearing and touch overlap in the lower frequency range. This apparent redundancy carries the advantages associated with the robustness afforded by multiple senses accessing a same physical quantity~\cite{Ernst-Bulthoff-04}, but are there other benefits? 

Comparisons made using discrimination or identification tasks are fraught with difficulties because hearing and touch excel in different perceptual domains. For example, hearing can track a single voice buried in a busy auditory landscape~\cite{Cherry-53,Bregman-90} and touch can instantly identify the materials of objects~\cite{Bergmann-Tiest-Kappers-06} even in the absence of thermal and textural cues~\cite{Gueorguiev-et-al-16}. 

The detection of weak stimuli, however, is a task where a fair comparison is possible. To this end, we used the physics of sound production to convert the tactile detection thresholds into sound pressure levels radiated by oscillating surfaces at a distance typical of an arm's length. Conversely, we expressed the auditory detection thresholds in terms of the vibration amplitudes of radiating surfaces. This method, which to our knowledge is novel, allowed us to quantify which modality is more effective at detecting an object according to its frequency of oscillation and to its size. Unexpectedly, we found that the objects around us that we can feel cannot be heard if they are small and if they move slowly.

\section*{Results}

The auditory threshold is defined to be the level at which a person can detect a sound half of the time during repeated trials~\cite{ISO226}. For pure tones produced by a sound source located in front of a listener and assuming plane wave propagation in free space, this threshold can be expressed by the sound pressure level (SPL) obtained from the acoustic pressure root-mean-square (RMS) value, $P_{\rm rms}$, using $\text{SPL}=20 \log_{10} (P_{\rm rms}/P_{\rm ref})$ where $P_{\rm ref}=2.0\times10^{-5}$~Pa is the reference pressure and where $P_{\rm rms}=P/\sqrt{2}$ if $P$ is the wave amplitude. 

Acoustic waves propagate in the auditory canal and set the eardrum into vibration. Vibrations are transmitted through the ossicles and the round window to produce auditory sensations. The knowledge of the impedance transfer function of this system would be required to resolve pressure thresholds in terms of displacements. 

\begin{figure}[h!]
\centering
\includegraphics[scale=1]{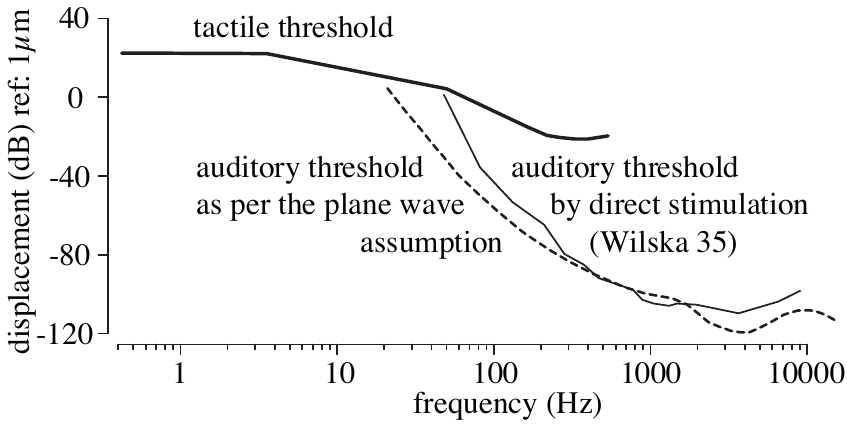}
\caption{Tactile detection threshold of the hand thenar emminence~\cite{Bolanowski-et-al-88}(thick line). Auditory threshold by direct excitation of the eardrum (thin line)~\cite{Wilska-35}. ISO226 auditory threshold~\cite{ISO226} expressed as displacement under the assumption of plane wave propagation~\eqref{PlaneWaveRatio} (dashed line).}
\label{BolanoVsWilska}
\end{figure}

To circumvent this problem, eighty years ago, Wilska developed a method to vibrate the eardrum through a rigid rod~\cite{Wilska-35}. This way, he could directly determine the auditory threshold in terms of vibration amplitude. Later, Geischeider used this result to compare auditory detection thresholds with tactile detection thresholds obtained by vibrating pistons impinging on the skin~\cite{Bolanowski-et-al-88}. The comparative diagram is reproduced in Fig.~\ref{BolanoVsWilska} showing that the eardrum is more sensitive than the skin for frequencies higher than 20--50~Hz. At its optimum, around 300~Hz, the tactile displacement threshold of the thenar eminence of the hand is indeed about 0.1~$\upmu$m, while hearing at the same frequency can detect displacements as small as 0.1~nm, limited only by quantum physics~\cite{Bialek-87}.

In air, the law of conservation of momentum links mechanical displacement to acoustic pressure. To a first order, if $\rho$ is the mass density of air, $\boldsymbol{u}$ the fluid particle displacement vector, and $p$ the acoustic pressure,
\begin{equation}\label{Newton2}
\rho \frac{\partial ^2 \boldsymbol{u}}{\partial t^2}=-\nabla p.
\end{equation}
In the case of a plane wave normal to the $x$ direction, the pressure, $p(x,t)$, has the form $P(\omega) e ^{j (\omega t - kx)}$, where $\omega = c k$ is the pulsation, if $c$ is the wave velocity, and $k$ the wavenumber. Substituting this pressure field in the wave equation~\eqref{Newton2} gives a particle displacement equal to $U_{x}(\omega)e^{j (\omega t - kx)}$ propagating along the $x$ direction. The ratio of acoustic pressure to particle displacement is then,
\begin{equation}\label{PlaneWaveRatio}
\frac{P}{U_{x}}(\omega)=\rho c j \omega =j Z \omega ,
\end{equation}
where $Z\approx413$~Pa$\cdot$s$\cdot$m$^{-1}$ is the impedance of air at 20~$^{\circ}$C. 

The plane wave model can be employed to describe the propagation of acoustic waves in the ear canal because its diameter ($\approx 8$~mm) is two orders of magnitude smaller than the wavelength ($\lambda\approx0.3$~m at one~kHz). The concha and the pinna, acting only at frequencies greater than 2~kHz~\cite{Shaw-74}, have no acoustic role to play in our results. The eardrum, having an impedance similar to that of air at low frequencies~\cite{Allen-86}, moves like the fluid particles impinging on it. Using \eqref{PlaneWaveRatio} to express the auditory threshold we found values expressed in terms of displacement that indeed give values that are remarkably close to those found empirically by Wilska, see Fig.~\ref{BolanoVsWilska}. 

Using these values to compare hearing and touch, however, is akin to considering a direct coupling of an oscillating surface with the eardrum. It does not account for the conditions in which the sensory organs normally operate. To make quantitative comparisons more relevant one should consider the radiation patterns produced by remote oscillating surfaces.

In their experiments, Bolanowski and predecessors used pistons vibrating the skin through an aperture~\cite{Verrillo-63,Bolanowski-et-al-88}. An oscillating piston surrounded by a fixed plane corresponds to the baffled piston model used in sound radiation problems. With this model, we can determine the sound pressure level radiated by a piston oscillating at the tactile detection threshold or the vibration amplitude of the piston required to produce an audible sound. 

The acoustic pressure produced on the axis of a baffled piston of radius, $a$, is~\cite[p.180]{Kinsler-et-al-00},
\begin{align*}
p(r,t)&=\rho c j \omega U_{\rm p}(\omega)\left(1-e^{-j k\left(\sqrt{r^2+a^2}-r\right)}\right) e ^{j (\omega t - kr)}\\
&=P(r,\omega) e ^{j (\omega t -k r)} ,
\end{align*}
where $U_{\rm p}(\omega)$ is the displacement amplitude of the piston and $P(r,\omega)$ is the pressure at a distance, $r$, and at pulsation, $\omega$. In the near field, delimited by $r<a^2/\lambda$ with $a>\lambda$, the ratio of pressure to displacement oscillates between zero and $2 Z \omega$. In the range from 20~Hz to 1~kHz, the wavelength is comprised between 16~m and 0.33~m. The near field region, where \eqref{PlaneWaveRatio} holds approximately, can therefore be neglected since pistons would have to be larger than the contact areas with fingers and hands. In the far field, $r/a \gg 1$ and $r/a>ka/2$, the expression for pressure amplitude simplifies to
\begin{equation}\label{EqPpiston}
P(r,\omega)\simeq\frac{1}{2}\frac{\rho a^2}{r}\omega ^2 U_{\rm p}.
\end{equation}
Thus, in the far field the ratio of pressure to displacement decreases with $1/r$ and we can compute the vibration amplitude needed to reach a given acoustic pressure amplitude, $P$, at a certain distance and frequency,
\begin{equation}\label{EqUpiston}
U_{\rm p}(r,\omega) = \frac{2 r}{\rho a^2\omega ^2} P.
\end{equation}

\begin{figure}[hhh]
\centering
\includegraphics[scale=1]{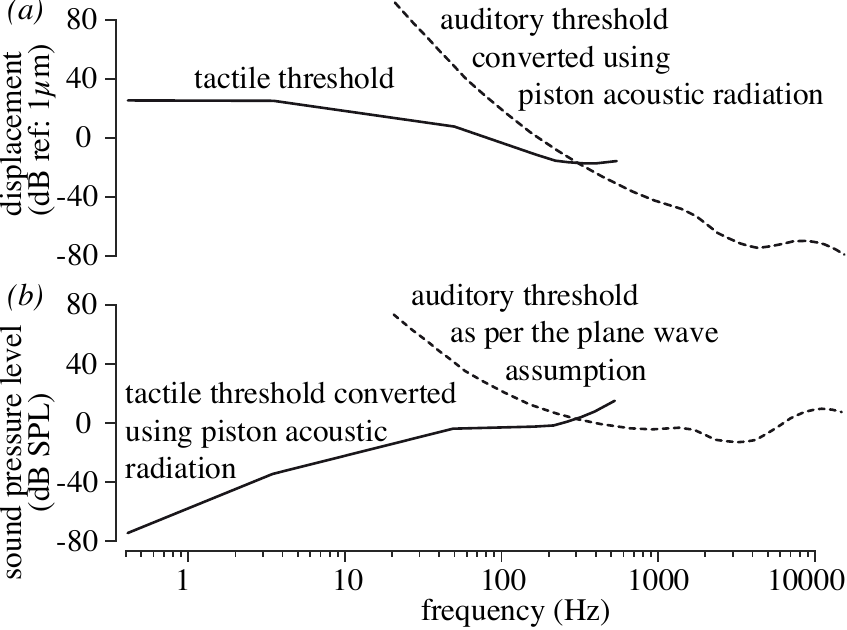}
\caption{({\bf a}) Tactile threshold~\cite{Bolanowski-et-al-88} (solid line) and piston oscillation amplitude required to reach the auditory threshold (dashed line). ({\bf b})~Auditory threshold~\cite{ISO226} (dashed line) and sound pressure level radiated at the same distance by the same piston oscillating at tactile detection threshold (solid line) ($a=9.6$~mm, $r=50$~cm, $\rho=1.20$~kg$\cdot$m$^{-3}$, $c=340~$m$\cdot$s$^{-1}$).}
\label{BolanoVsIsoVsSPL}
\end{figure}

Figure \ref{BolanoVsIsoVsSPL}a shows the vibration amplitude required to produce a sound audible from a distance of 50~cm---typical of the distance of the hand to the ear---by listening to the piston of radius 9.6~mm used by Bolanowski. The tactile detection threshold found using this same source is also shown. 

Using the same data and model, Fig.~\ref{BolanoVsIsoVsSPL}b shows the sound pressure level produced by the same piston oscillating at the tactile detection threshold and compares it with the auditory detection threshold. Over most of the tactile sensitivity range, the vibration of a piston can be felt but produces no audible sound.

The threshold curves now paint a very different picture from that of Fig.~\ref{BolanoVsWilska}. The tactile detection threshold is below the auditory threshold for all frequencies below $300$~Hz. Touch is indeed more sensitive than hearing in a large range of frequencies. 

Acoustic pressure is affected by the size of the piston and by its distance to the listener according to expression~\eqref{EqPpiston}. Since acoustic pressure changes with $1/r$, the tactile threshold represented in Fig.~\ref{BolanoVsIsoVsSPL}b should be increased by 6~dB when the distance to the vibrator is halved. Verrillo~\cite{Verrillo-63} measured how the tactile detection threshold varied with the piston area and noticed that the tactile detection threshold decreased by 3~dB per doubling piston area, that is, proportionally to $1/\sqrt{S}$. Equation \eqref{EqUpiston} shows that the displacement needed to reach a given acoustic pressure changes as $1/S$. These scaling laws proclaim that, as an oscillating surface decreases in size, the tactile sensibility decreases at a slower rate than the radiated sound pressure level providing even more advantage to touch when the objects to be detected are small.

\begin{figure}[!h]
\centering
\includegraphics[scale=1]{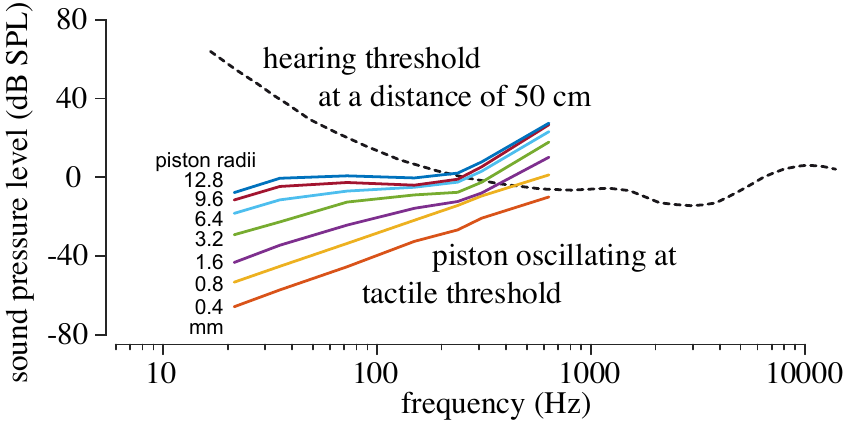}
\caption{Auditory threshold and sound pressure level radiated at a distance of 50~cm by various pistons oscillating at the tactile detection threshold~\cite{Verrillo-63}.}
\label{VerilloVsIso}
\end{figure}

This trend can be seen more clearly in Fig.~\ref{VerilloVsIso} where the sound pressure level calculated from the tactile detection threshold values using various piston radii are reported together with the auditory detection threshold. Small objects, indeed, produce inaudible sounds over the entire tactile sensitivity range. Only larger objects produce audible sounds as well as tangible vibrations.

To quantify these findings more precisely, Table~\ref{SPLdiff} collects the differences in sound pressure level and the auditory threshold for an oscillating piston at a distance of 50~cm. Positive values indicate the production of audible sound at the corresponding tactile detection threshold. To use this table for other listening distances, an increment of +6 dB  per halved distances should be added to these values. 

Lastly, for oscillating pistons, movement amplitude, frequency, and speed obey the scaling law,
\begin{equation*}
|\dot U_{\rm p}| \propto \omega |U_{\rm p}|,
\end{equation*}
which, for a given movement amplitude, relates frequency to speed.

\begin{table}[hhh]
\caption{Difference in sound pressure level (dB SPL) relative to the auditory threshold for a source at 50~cm from the ear. Negative values indicate felt objects that cannot be heard.}
\label{SPLdiff}
\begin{center}
\setlength{\tabcolsep}{6pt}
\begin{tabular}{r|rrrrrrr}
\multicolumn{1}{r}{}&\multicolumn{6}{c}{frequencies (Hz)} \\
\multicolumn{1}{r}{} & 40 & 80 & 160 & 250 & 320 & 640 & \\
\cline{2-7}
\rule{0pt}{10pt}
\multirow{7}{*}{\begin{turn}{90}piston radii (mm)\end{turn}}
12.8 & -42 & -21 & -8 & {\bf 0} & {\bf 9} & {\bf 34} \\
9.6 & -46 & -25 & -12 & -3 & {\bf 6} & {\bf 33} \\
6.4 & -53 & -29 & -13 & -5 & {\bf 4} & {\bf 30} \\
3.2 & -64 & -35 & -17 & -10 & -2 & {\bf 24} \\
1.6 & -77 & -47 & -24 & -15 & -7 & {\bf 16} \\
0.8 & -88 & -57 & -30 & -17 & -9 & {\bf 7} \\
0.4 & -100 & -69 & -41 & - 29 & -20 & - 4 
\end{tabular}
\end{center}
\end{table}

\section*{Discussion}

Under evolutionary pressure, our senses have probably adapted to take advantage of the ambient physics. Objects around us cover many orders of magnitudes in term of sizes, distances, and movement speeds. Having found that hearing is not always more apt than touch at detecting the small displacements of objects suggests that hearing and touch have co-evolved to extend the range of detectable of objects.

When objects are small and oscillate slowly touch is the only sensory option for their detection because the acoustic energy that they radiate is small. It is in the high frequency range where the objects oscillate rapidly, that hearing is more effective than touch: mosquitoes can be heard and felt, but ants can only be felt! 

Of course, the relative sensitivity of touch and hearing is also affected by many other factors such as the location of stimulation on the body, the distance between the source and the listener, or the orientation of the acoustic source with respect to the listener; and we only considered acoustic sources realized by surfaces with uniform out-of-plane motion. In particular, the results would differ in the case of vibrations propagating in a radiating surface. The trend that we described would nevertheless hold broadly and should be accounted for the design of tactile stimulators and audio devices. 

\bigskip

\noindent\textbf{Acknowledgments.} The writing of this article benefited from insightful comments from Ophelia Deroy, David Gueorguiev, David Alais, Ana Tajadura-Jim\'enez, and from help from Jos\'e Lozada. 

\noindent\textbf{Authors' contributions.} CH and VH performed the research and wrote the paper. 

\noindent\textbf{Funding.} Work supported by the European Research Council (FP7 Program) ERC Advanced Grant (PATCH) to VH (No. 247300).

\noindent\textbf{Competing interests.} The authors declare no competing interests.

\bigskip

\bibliography{TouchHearing}

\end{document}